# A DATASET FOR MEASURING READING LEVELS IN INDIA AT SCALE

*Dolly Agarwal[1], Jayant Gupchup[2], Nishant Baghel[1]*

[1]Pratham Education Foundation, [2]Pratham Volunteer[*]
[1]{dolly.agarwal, nishant.baghel}@pratham.org, [2]gupchup@gmail.com

## ABSTRACT[1]

One out of four children in India are leaving grade eight without basic reading skills. Measuring the reading levels in a vast country like India poses significant hurdles. Recent advances in machine learning opens up the possibility of automating this task. However, the datasets of children's speech are not only rare but are primarily in English. To solve this assessment problem and advance deep learning research in regional Indian languages, we present the ASER dataset of children in the age group of 6-14. The dataset consists of 5,301 subjects generating 81,330 labeled audio clips in Hindi, Marathi and English. These labels represent expert opinions on the child's ability to read at a specified level. Using this dataset, we built a simple ASR-based classifier. Early results indicate that we can achieve a prediction accuracy of 86% for the English language. Considering the ASER survey spans half a million subjects, this dataset can grow to those scales.

## 1. INTRODUCTION

According to UNESCO Institute for Statistics (UIS)[1], more than 617 million children and adolescents are not achieving minimum proficiency levels (MPLs) in reading and mathematics. Similar results have been reported in the Programme for International Student Assessment (PISA[2]), conducted in the states of Himachal Pradesh and Tamil Nadu in India.

A majority of children need immediate help in acquiring foundational skills in literacy, but estimating the reading levels of individual children at scale becomes a key challenge. The challenge is significant in a country like India where the pupil teacher ratio is more than 30 [3].

ASER is a household-based standardized survey of children's schooling and learning status. The ASER survey is conducted by Pratham volunteers from local partner organizations in many districts in India [4, 5]. Since 2005, Pratham volunteers are trained intensively to collect data on reading levels on pen and paper [5]. An external evaluation of ASER conducted in 2013-14 calculated that the ASER survey costs a little over Rs 100 per child (approximately U.S. $1.40). This cost albeit lower compared to other large-scale learning assessments is notable because ASER involves assessing about 700,000 children every year. In addition, depending on the age and ability of the child, the assessment of reading and arithmetic takes an average of about ten minutes per child. An automated system can help save the time and cost involved every year in training the volunteers and conducting the test.

Recent research [6][7][8][9] in deep learning and speech recognition technology to assess English reading proficiency has made it possible to develop computer-based reading evaluators that listen to students, assess their performance, and provide a score or immediate customized feedback. In comparison, there has been far less work done in similar areas for children speaking Indian languages. Moreover, the dataset used for current solutions are limited [6][9]. Unlike adult speech corpora, there are less speech data of children [10]. Most children's speech databases which are available are either in English or some European languages. Databases of children's speech are required in order to improve ASR for children. The differences in languages, pronunciation and accent across India pose many challenges in the development of accurate evaluation tools. Furthermore, a child speech corpus can be used in the development of computer-assisted language learning systems, reading tutors, tools for foreign language learning, computer games etc. which are less relevant to adults. Thus, databases of children's speech are required in order to investigate ASR for children

Targeting to solve this problem, the first version of the ASER dataset was collected for Hindi, Marathi and English using an Android app over a period of 6 months from three states of India - Uttar Pradesh, Maharashtra and Rajasthan [11]. The app captures the reading proficiency level of the child along with the number of mistakes made at each level. Representative screenshots of the app are shown in Figure 1. The dataset includes audio files of children reading out different levels of text using omnidirectional condenser microphone and recorded in 16 kHz sampling frequency. Based on this initial dataset, a simple classifier was built using features extracted from commercially available speech-to-text (STT) services. This classifier is able to achieve an accuracy of up to 86%. To the best of our knowledge, this is the first body of work that has curated datasets in multiple Indian languages with the purpose of assessing the reading levels of children. Therefore, the dataset that we present is unique in itself because we not only introduce a substantial size of children's speech but we also bring in speech data in two Indian languages (Hindi and Marathi).

---

[1] Data and Source code is available at
https://github.com/PrathamOrg/ASER-Dataset.git
[*] Currently affiliated with Microsoft. This work was done as a Pratham volunteer and is not connected with Microsoft.

The solutions obtained using this dataset can help Pratham automate the process of identifying students with reading difficulty, and for teachers to adopt the "Teaching at the Right Level" (TaRL[12]) approach. Same logic and models can be extended to other TaRL inspired countries - Botswana, Côte d'Ivoire, Ghana, India, Kenya, Madagascar, Mozambique, Niger, Nigeria, Uganda, Zambia by generating relevant dataset. This will extend the reach to over 60 million children. Additionally, this open source dataset will also benefit the ecosystem of academia and researchers working in speech and language, Speech-to-text for Indian languages and prosodic skills evaluation. The models developed in turn will enable tech companies working on adaptive learning tools/products to cater to the needs of Indian education system. The main contributions of this paper are:
- A novel methodology to collect data at scale for solving the problem of automating reading levels of children in Indian languages.
- A dataset of 5,301 children – 81,330 labeled audio clips and 123 hours of speech representing children in the age group of 6-14 collected in two Indian languages and English.
- A baseline solution for assessing the reading levels in an automated fashion that achieves an accuracy of 86% and a precision of 90% at the 80% recall level using commercially available STT services.

The rest of the paper is organized as follows: Section 2 details the related work. In Section 3, we present the methodology used to collect data. Section 4 describes the dataset and the baseline solution, and finally, in Section 5 we conclude.

## 2. RELATED WORK

A number of speech databases are available for the English language. Some notable ones include Blizzard challenge [13], Voxforge [14], Mozilla common voice [15]. A single speaker dataset with 10 hours of audio is available in the Japanese language [16]. Tundra is a comprehensive dataset with more than 14 languages built from audiobooks [17]. Perhaps the closest dataset to the one presented in this paper is the Low Resource Automatic Speech Recognition Challenge for Indian Languages [18]. It has over 150 hours of Tamil, Telugu and Gujarati languages. Google has a dataset with over 11,000 videos and 32 hours of children's speech [19]. To the best of our knowledge, this is the first comprehensive dataset made available for assessing reading levels among children in Indian languages.

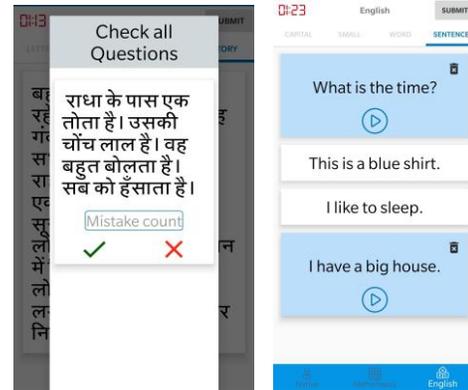

Figure 1: UI for Native Language (left) and English (right)

The automated assessment of reading levels has been a widely researched topic in recent years. Systems have been built using measures related to pronunciation, fluency, and reading accuracy, words per minute count, and have achieved high correlations with human judgments of English proficiency, e.g. [6, 7, 8, 9]. Sabu et al. have employed Automated Speech Recognition (ASR) and prosody modeling for evaluating the reading levels of children [9]. In their work, word decoding accuracies and prosody attributes like phrasing and prominence are considered for assessment. The relationship between different acoustic features computed from the speech signal and the perceived quality is also discussed. Hacker et al. propose to assess a learner's pronunciation by developing a large set of nearly 200 pronunciation and prosodic features [20]. Using this approach, pronunciation scoring is regarded as a classification task in high-dimensional feature space.

A number of adaptive learning products have implemented an automated system to assess the reading levels of children. Group Reading Assessment and Diagnostic Evaluation Online (GRADE Online [21]) by Pearson assessment is a product available for schools in the US at a premium fee. GRADE™ is a diagnostic reading test that determines what developmental skills PreK–12 students have mastered and where students need instruction or intervention. NWEA, the not-for-profit creator of assessment solutions, introduced MAP Reading Fluency [22] in 2018, the first computer adaptive, automatically scored K-3 oral reading assessment that uses Speech Recognition, Automatic Scoring and Computer Adaptive Technology. TruAccent speech-recognition engine by Rosetta Stone Inc. [23] helps fine tuning pronunciation in 24 languages including Hindi and is available for businesses and schools.

| FIELD | DESCRIPTION | EXAMPLE |
|---|---|---|
| ageGroup | Age group of the child | 5-7 years |
| studClass | School Grade | 1 |
| Date | Test Date and Time | Thu Jul 04 15:15:07 IST 2019 |
| NativeProficiency | Highest Level that the child could speak comfortably as reported by the assessor | Paragraph |
| EnglishProficiency | Highest Level that the child could speak comfortably as reported by the assessor | Sentence |

Table 1: Subject specific details captured in the json

| FIELD | DESCRIPTION | EXAMPLE |
|---|---|---|
| IsCorrect | Test label. True if test is read correctly; otherwise false | True |
| NoOfMistakes | Number of mistakes done while reading test | 1 |
| que_id | Id of question | HI_S1_S_2 |
| que_text | Text displayed to read | I like to read. |
| recordingName | Name of audio clip | HI_S1_S_2.mp3 |

Table 2: Test specific details of one subject captured in json

## 3. METHODOLOGY

### 3.1. Data Collection

An Android app was designed to collect human evaluated training data [11]. The app was used to collect data from three states of India - Maharashtra, Uttar Pradesh and Rajasthan. Figure 1 are screenshots of the designed UI of the app. While conducting the test, the audio clips were recorded for each section and the assessor marked the correctness of every question. After each section, the proficiency level of the child was marked by the assessor. The details of the test were recorded in a json file as described in Table 1 and 2. After the test, the assessor pushes the data which uploads it to Firebase Cloud Storage. Every Pratham staff goes through a rigorous training before they can start collecting data. It is worth noting that since ASER is a standardized test, the vocabulary size is finite. This simplifies the problem considerably and standardized utterances ensure that there is no PII in the entire dataset.

### 3.2. Noise Reduction Effort

The initial round of data was gathered using an 8 kHz sampling rate and an onboard microphone device. The audio quality needed improvement; hence, we discarded those samples. We changed the sampling rate from 8 kHz to 16 kHz. We also employed an omnidirectional condenser microphone for recordings. These changes helped improve the audio quality significantly.

### 3.3. Validation of labels and assessors

A web portal is designed to ensure the quality of collected samples and labels. 5 ASER experts are given random samples from the collection. They listen to the recorded audios and verify the labels marked by the assessors.

## 4. DATASET AND RESULTS

### 4.1. The Dataset

There are 5301 sample folders: a sample folder contains multiple audio files and its corresponding json file. The distribution of the sample types based on the language, proficiency level and age group wise is shown in Table 3 and Figure 2.

| Hindi (3959 subjects) | | | | | |
|---|---|---|---|---|---|
| | Story | Para | Word | Letter | Total |
| No. of distinct Samples | 4 | 8 | 41 | 17 | 70 |
| No. of audio clips | 2244 | 3290 | 4669 | 5667 | 15870 |
| Duration (hrs) | 28.30 | 14.81 | 5.65 | 4.92 | 53.68 |
| Marathi (1342 subjects) | | | | | |
| | Story | Para | Word | Letter | Total |
| No. of distinct Samples | 4 | 8 | 36 | 23 | 71 |
| No. of audio clips | 860 | 1225 | 1307 | 1106 | 4498 |
| Duration (hrs) | 14.68 | 5.40 | 1.45 | 0.99 | 22.52 |
| English (5047 subjects) | | | | | |
| | Sentence | Word | CL | SL | Total |
| No. of distinct Samples | 22 | 28 | 24 | 23 | 97 |
| No. of audio clips | 6224 | 14611 | 22186 | 17941 | 60962 |
| Duration (hrs) | 9.80 | 13.00 | 14.14 | 10.34 | 47.28 |

Table 3: Distribution of samples – language & level wise

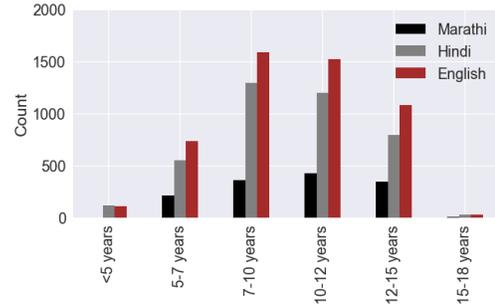

Figure 2: Distribution of samples age-group wise

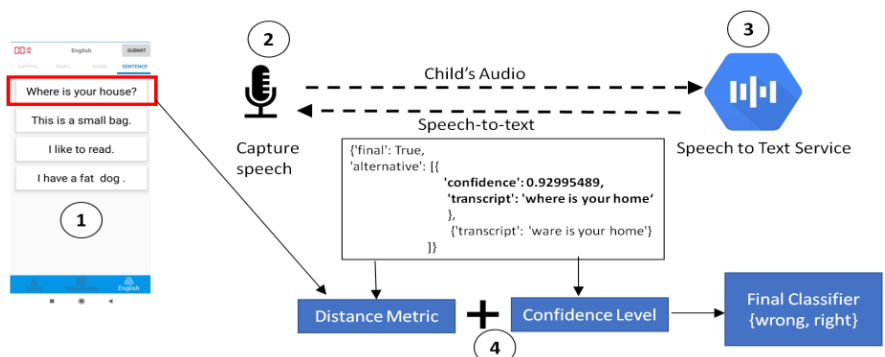
Figure 3: The end-to-end architecture of the classifier design for sentence level in English.

*4.1.1. The question samples*
For each language (Hindi and Marathi), there are 4 question sample sets. Each question sample set consists of 2 sections - Native Language and English tools. Each section contains few samples[2] of respective levels: Story, Paragraph, Words and Letters for Native Language; Sentence, Words, Capital letters and Small letters for English. The words in sentence, paragraph and story are different from the words in word section.

*4.1.2. The audio files*
The utterances of a child reading out different levels of texts are recorded and saved as mp3 audio files. There are multiple audio files for one child. For reading section the audio files can be one paragraph, one story, 5 words (separate audio files for each), 5 letters (separate audio files for each). Similarly, for English section, the audio files can be of 5 capital letters, 5 small letters, 5 words and 4 sentences.

**4.2. Baseline Solution**

Our team has used this dataset to come up with a baseline solution for English sentences. The overall architecture of the solution is captured in Figure 3.

The captured audio of the child's speech is sent to a commercial speech-to-text service. We tested this solution using Google's speech-to-text service and Microsoft's Azure's cognitive service and the results were comparable in terms of statistical significance [24, 25]. The response and confidence level from the service are extracted. Since the original test phrase is known, we can use the transcribed text to compute a distance metric such as edit distance. The edit distance and confidence level are fed to train a simple random forest classifier to detect if a child is at the sentence level or not. The classifier is trained using the scikit-learn python package [26]. We balanced the classes and split the original data into train/test splits. We are able to obtain an accuracy of 86%. In our application, we have a bias towards precision. This is because a false positive implies that a child could be left behind. In comparison, a false negative is relatively less costly since the child will be slightly ahead of his/her cohort. Using this method, at a recall level of 80%, we are able to achieve precision of 90%.

It is worth mentioning that while the current solution can be deployed in the field, the target accuracy level for such an automated solution in the field is greater than 95%. This level of performance will ensure efficient utilization of resources.

**5. CONCLUSION**

In this paper we have provided a fully annotated dataset to assess the reading levels in India and facilitate research in speech processing. In order to establish a baseline solution, we have developed a simple random forest classifier using features obtained from commercial STT services. This approach is able to achieve a classification accuracy of 86% for the English sentence level. We expect the accuracy to improve as we gather more data. In the next phase of this work, we will apply this methodology to Hindi and Marathi as well. With Pratham's reach of 4,200 villages in India and 1,000 trained volunteers, we can scale the dataset easily to go up to half a million samples in 11 regional languages. Finally, we would like to emphasize the impact of coming up with innovative solutions to solve this practical and real-world problem. An automated solution for assessing the reading levels in India will result in saving ~1 million dollars and 10 minutes time per assessment. We hope that this dataset mutually benefits the speech research community and solves a very practical problem for society at scale.

**6. ACKNOWLEDGEMENTS**

We would like to thank Google for funding the full labeled data collection. Yaran Fan, Joyce Fang, Aditya Khant and Kshitij Gupta from Microsoft for providing us with a baseline solution in a non-profit hackathon (hack-for-good). We thank Impact Analytics for the initial design of the collection app. Pravin Kate, Kumar Ashwarya and Ayesha Selwyn from the Pratham team improved the app and data collection process. Lastly, we thank Prof. Preeti Rao, IIT Bombay for her suggestions for improving the quality of audio recordings.

---

[2] Complete Vocabulary details is available at:
https://github.com/PrathamOrg/ASER-Dataset.git